# Size effects and depolarization field influence on the phase diagrams of cylindrical ferroelectric nanoparticles.

Anna N.Morozovska[*],

[*]V. Lashkaryov Institute of Semiconductor Physics, NAS of Ukraine,

41, pr. Nauki, 03028 Kiev, Ukraine, morozo@mail.i.com.ua

Eugene A. Eliseev[**], Maya D.Glinchuk[**]

[**]Institute for Problems of Materials Science, NAS of Ukraine,

Krjijanovskogo 3, 03142 Kiev, Ukraine, glin@materials.kiev.ua

**Abstract**

Ferroelectric nanoparticles of different shape and their nanocomposites are actively studied in modern physics. Because of their applications in many fields of nanotechnology, the size effects and the possible disappearance of ferroelectricity at a critical particle volume attract a growing scientific interest. In this paper we study the size effects of the cylindrical nanoparticle phase diagrams allowing for effective surface tension and depolarization field influence. The Euler-Lagrange equations were solved by direct variational method. The approximate analytical expression for the paraelectric-ferroelectric transition temperature dependence on nanoparticle sizes, polarization gradient coefficient, extrapolation length, effective surface tension and electrostriction coefficient was derived. It was shown that the transition temperature could be higher than the one of the bulk material for nanorods and nanowires in contrast to nanodisks, where the decrease takes place. The critical sizes and volume of ferroelectric-paraelectric phase transition are calculated. We proved that among all cylindrical shapes a nanobar reveals the minimal critical volume. We predicted the enhancement of ferroelectric properties in nanorods and nanowires. Obtained results explain the observed ferroelectricity enhancement in nanorods and could be very useful for elaboration of modern nanocomposites with perfect polar properties.

PACS: 77.80.-e, 77.84.Dy, 68.03.Cd, 68.35.Gy

## 1. Introduction

Ferroelectric nanoparticles of different shape are actively studied in nano-physics and nano-technology. Because of miniaturization of devices based on these materials, the study of ferroelectric properties size dependence and the possible disappearance of ferroelectricity at a finite critical volume attract a high scientific interest.

The ferroelectric phase was studied in ferroelectric nanowires, nanotubes and nanorods [1], [2], [3], [4], [5]. It is appeared that nanorods and nanowires posses such polar properties as remnant polarization and piezoelectric hysteresis [1], [2], [5]. Moreover, the confined geometry does not destroy ferroelectric phase as predicted for spherical particles [6], [7] and observed experimentally [8], [9], [10], but sometimes the noticeable enhancement of ferroelectric properties appears in nano-cylinders [1], [2], [3], [4], [5], [11]

Yadlovker and Berger [1] reported about the spontaneous polarization enhancement up to 0.25-2$\mu C/cm^2$ and ferroelectric phase conservation in Rochelle salt (RS) ($NaKC_4H_4O_6 \cdot 4H_2O$) nanorods (radius about 30nm and height 500nm) up to the material decomposition temperature $55^0C$ that is about $30^0C$ higher than the one of bulk-size crystals.

Mishina *et al* [11] revealed that ferroelectric phase exists in $PbZr_{0.52}Ti_{0.48}O_3$ (PZT) nanorods with diameter less than 10-20nm. However, earlier Mishra *et al* [9] demonstrated that the ceramics prepared from the powders of $PbZr_{0.52}Ti_{0.48}O_3$ with size about 100 nm had a pseudo-cubic symmetry, but might exhibit a hump in the temperature variation of dielectric constant. So, it seems that the critical size for the PZT nanorod (if any) is about 10 times smaller than the one for the nanosphere.

Geneste *et al* [2] studied the size dependence of the ferroelectric properties of $BaTiO_3$ (BT) nanowires from the first principles. They showed that the ferroelectric distortion along the wire axis disappears below a critical size of about 1.2nm. Note, that $BaTiO_3$ spherical nanoparticles have the much larger critical size of about several tens nm [10], [12].

Morrison *et al* [5] demonstrated that ultra-small $PbZr_{0.52}Ti_{0.48}O_3$ nanorods and nanotubes (radius 20-30nm, length $50\mu m$) possesses rectangular shape of the piezoelectric hysteresis loop with effective remnant piezoelectric coefficient value compatible with the ones typical for PZT films [13]. This fact unambiguously speaks in favor of spontaneous polarization existence. Also the authors demonstrated that the ferroelectric properties of the free $BaTiO_3$ nanotubes are perfect.

Thus, at the first glance recent experimental results contradict the generally accepted viewpoint that the ferroelectric properties disappear under the system volume decreases below the critical one [14]. Actually the aforementioned facts proved that the shape of nanoparticles (e.g. the spherical or cylindrical one) essentially influences on the minimal sizes necessary for the ferroelectricity conservation [2] possibly owing to the different depolarization field and mechanical boundary conditions [15], [16]. Immediately, one should ask the principal question: What nanoparticle shape posses the minimal critical volume and allows ferroelectricity conservation at higher temperatures? Could the answer be predicted theoretically?

In theoretical papers [6], [17] the special attention was paid to size effects of nanoparticles, but finite cylinders were not considered. It is well known that depolarization field exists in the majority of confined ferroelectric systems [18] (including the cylindrical nanoparticles) and could cause the aforementioned size-induced ferroelectricity disappearance in insulator single-domain films and ellipsoidal particles [19], [20], [21]. As a result of ferroelectric properties degradation, the phase transition temperature in spherical nanoparticles is significantly lower then the bulk one for most of the cases [6], [22], [21], [8]. Since the depolarization field value depends on the shape of a particle, the enhancement of ferroelectricity can be expected for the ones with smaller depolarization field.

In this paper we study the size effects, surface tension and depolarization field influence on the cylindrical nanoparticles properties. We suppose that a nanoparticle surface is covered with a charged layer consisted of the free carriers adsorbed in the ambient conditions (e.g. air with definite humidity or pores filled with a precursor water solution). For instance a thin water layer condensates on the polar oxide surface in the air with humidity 20-50% [23] The surface charges screen the surrounding medium from the nanoparticle electric field [14], but the depolarization field inside the particle is caused by inhomogeneous polarization distribution. As a matter of fact we calculated the depolarization field inside a cylindrical nanoparticle under the short-circuit conditions proposed by Kretschmer and Binder [19].

For the description of nanodisks, nanorods and nanowires ferroelectric properties we used the Euler-Lagrange equations solved by means of a direct variational method [20]. The approximate analytical expression for paraelectric-ferroelectric transition temperature dependence on the nanoparticle sizes, extrapolation length, effective surface tension coefficient, polarization gradient and electrostriction coupling coefficients *etc* was derived. We obtained, that the possible reason of the polar properties enhancement in confined ferroelectric nanowires and nanorods is the effective surface pressure coupled with polarization via the electrostriction effect and the decrease of depolarization field value occurred in prolate cylindrical particles.

## 2. Free energy with renormalized coefficients

Let us consider ferroelectric cylindrical nanoparticle with radius $R$, height $h$ and polarization $P_Z(\rho, z)$ oriented along $z$–axes. Hereinafter $V = \pi R^2 h$ is the particle volume, the polarization distribution $P_Z(\rho, z)$ is axisymmetric. The external electric field is $\mathbf{E} = (0, 0, E_0)$ (see Fig. 1).

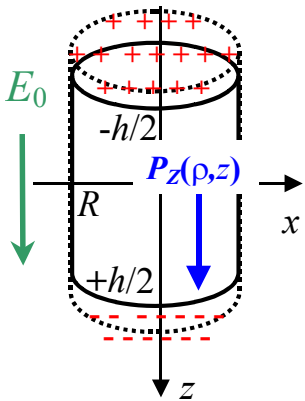

FIG. 1. (Color online) The geometry of cylindrical particle.

The coupled equations for the polarization calculations can be obtained by the variation on polarization of the free energy functional $\Delta G = \Delta G_V + \Delta G_S$ consisted from the bulk part $\Delta G_V$ and the surface one $\Delta G_S$ (see e.g. Refs. 15, 16). The bulk part $\Delta G_V$ acquires the form:

$$\Delta G_V = 2\pi \int_{-h/2}^{h/2} dz \int_0^R \rho d\rho \left( \begin{array}{l} \dfrac{\alpha_R(T)}{2} P_Z^2(\rho,z) + \dfrac{\beta_Z}{4} P_Z^4(\rho,z) + \dfrac{\gamma_Z}{6} P_Z^6(\rho,z) + \\ + \dfrac{\delta}{2}(\nabla P_Z(\rho,z))^2 - P_Z(\rho,z)\left( E_0 + \dfrac{E_Z^d(\rho,z)}{2} \right) \end{array} \right) \quad (1)$$

Material coefficients $\delta > 0$, $\gamma_Z > 0$, while $\beta < 0$ for the first order phase transitions or $\beta > 0$ for the second order ones. The coefficient $\alpha_R(T)$ in Eq.(1) should be renormalized by the external stress (see e.g. Ref. 22). In Appendix A we study the influence of the effective surface tension on a cylindrical particle and derived the expression for $\alpha_R(T)$:

$$\alpha_R(T,R) = \alpha_T(T - T_C) + 2Q_{12}\frac{\mu}{R} \quad . \quad (2)$$

Here parameters $T_C$, $\alpha_T$ and $Q_{12}$ are respectively Curie temperature, inverse Curie constant, and electrostriction coefficient regarded known for the bulk material. The parameter $\mu$ is the effective surface tension coefficient between the nanoparticle and interface [24], [10].

Note, that the renormalization of coefficient $\alpha_R = \alpha + 2Q_{12}p$ for a cylindrical nanoparticle differs from the one $\alpha_R = \alpha + (Q_{11} + 2Q_{12})p$ obtained for a spherical particle [22] ($p$ is the pressure applied to the particle surface). Both results are clear owing to the fact that stresses $\sigma_1 = \sigma_2 = -p$ and $\sigma_3 = 0$ for a cylinder, whereas $\sigma_1 = \sigma_2 = \sigma_3 = -p$ for a sphere. Also we do not take into account possible stress relaxation caused by dislocations. This approach used by many authors is valid for the small enough particle sizes [25].

The calculation of the depolarization electric field inside the finite cylindrical nanoparticle appeared to be more complex then for a spherical one. The exact expression for depolarization field $\mathbf{E}^d(\rho,z)$ inside the cylindrical nanoparticle covered with screening charges is derived in Appendix B (see Eq.(B.6)). Hereinafter we use its Pade approximation:

$$\begin{cases} E_Z^d(\rho,z) \approx \eta(R,h) \cdot (\langle P_Z \rangle - P_Z(\rho,z)), \\ \eta(R,h) = \dfrac{4\pi}{1 + (h/2R)^2} \end{cases} \quad (3)$$

The angular brackets are the spatial averaging on the particle volume $V$, e.g. $\langle P_Z \rangle \equiv \dfrac{2\pi}{V} \int_{-h/2}^{h/2} dz \int_0^R \rho d\rho P_Z(\rho,z)$. The function $\eta \sim (2R/h)^2 \ll 1$ for the prolate cylinder with $R \ll h$ [18], whereas $\eta \to 4\pi$ for the oblate cylinder with $R \gg h$ [19]. It should be noted that the depolarization field is absent outside the particles in the framework of our model. Therefore the interaction of such nanoparticles is practically absent due to the screening and they can be considered as the assembly of independent particles.

The surface part of the polarization-dependent free energy $\Delta G_S$ is supposed proportional to square of polarization on the particle surface $S$, namely $\Delta G_S = \frac{\delta}{2} \int_S \frac{ds}{\lambda} P_S^2$ ($\lambda$ is the extrapolation length [6], [17]). A cylindrical nanoparticle has upper and bottom surfaces $z = h/2$, $z = -h/2$ and a sidewall $\rho = R$ (see Fig. 1), so its surface energy $\Delta G_S$ acquires the form:

$$\Delta G_S = \delta \left( \int_0^R \frac{2\pi\rho}{\lambda_b} d\rho \left( P_Z^2\left(\rho, z = \frac{h}{2}\right) + P_Z^2\left(\rho, z = -\frac{h}{2}\right) \right) + \int_{-h/2}^{h/2} \frac{2\pi R dz}{\lambda_S} P_Z^2(\rho = R, z) \right). \tag{4}$$

We introduced longitudinal and lateral extrapolation lengths $\lambda_b \neq \lambda_S$ in Eq.(4). Hereinafter we regard these extrapolation lengths positive.

Variation of the free energy expressions (1) + (4) yields the following Euler-Lagrange equations with the boundary conditions on the cylinder faces $z = \pm h/2$, and the sidewall surface $\rho = R$ (see e.g. Refs. 6, 19, 20):

$$\begin{cases} \alpha_R P_Z(\rho, z) + \beta P_Z^3(\rho, z) + \gamma P_Z^5(\rho, z) - \delta \left( \frac{\partial^2 P_Z(\rho, z)}{\partial z^2} + \frac{1}{\rho} \frac{\partial}{\partial \rho} \rho \frac{\partial}{\partial \rho} P_Z(\rho, z) \right) = E_0 + E_Z^d(\rho, z), \\ \left( P_Z + \lambda_b \frac{dP_Z}{dz} \right)\bigg|_{z=h/2} = 0, \quad \left( P_Z - \lambda_b \frac{dP_Z}{dz} \right)\bigg|_{z=-h/2} = 0, \quad \left( P_Z + \lambda_S \frac{dP_Z}{d\rho} \right)\bigg|_{\rho=R} = 0, \end{cases} \tag{5}$$

Let us find the approximate solution of the nonlinear Eq.(5) by using the direct variational method as proposed earlier [20]. Firstly we solved the linearized Eq.(5):

$$P_Z(\rho, z) = \sum_{n=1}^{\infty} \frac{2J_0(k_n \rho/R)}{k_n J_1(k_n)} \left( \frac{E_0 + \eta \langle P_Z \rangle}{\alpha_R + \eta + \delta(k_n/R)^2} (1 - \varphi_n(z)) \right) \tag{6}$$

$$\varphi_n(z) = \frac{ch(\xi_n z)}{ch(\xi_n h/2) + \lambda_b \xi_n sh(\xi_n h/2)}, \quad \begin{cases} \xi_n = \sqrt{(k_n/R)^2 + (\alpha_R + \eta)/\delta} \\ J_0(k_n) - (\lambda_S/R) k_n J_1(k_n) = 0 \end{cases} \tag{7}$$

Here $J_0(k_n)$ and $J_1(k_n)$ are Bessel functions of the zero and first orders respectively. In general case the roots $k_n$ depend over the ratio $(\lambda_S/R)$ in accordance with Eq.(7). Under the condition $(\lambda_S/R) \ll 1$ one obtains that $J_0(k_n) \approx 0$. For the case we used the Bessel functions norm $\langle J_0(k_n \rho/R) J_0(k_m \rho/R) \rangle = \delta_{nm} J_1^2(k_n)$ and equality $\langle J_0(k_m \rho/R) \rangle = 2J_1(k_m)/k_m$ in Eq.(7) [26].

The average polarization induced by the external electric field in paraelectric phase and dielectric permittivity acquire the form:

$$\langle P_Z \rangle = \frac{\sum_{n=1}^{\infty} \frac{4}{k_n^2} \left( \frac{E_0(1-\Phi_n)}{\alpha_R + \eta + \delta(k_n/R)^2} \right)}{1 - \eta \sum_{n=1}^{\infty} \left( \frac{4}{k_n^2} \cdot \frac{1-\Phi_n}{\alpha_R + \eta + \delta(k_n/R)^2} \right)}, \quad (8)$$

$$\langle \varepsilon_{zz} \rangle = \frac{d\langle P_Z \rangle}{dE_0} = \frac{4\pi \sum_{n=1}^{\infty} \frac{4}{k_n^2} \frac{(1-\Phi_n)}{\alpha_R + \eta + \delta(k_n/R)^2}}{1 - \eta \sum_{n=1}^{\infty} \left( \frac{4}{k_n^2} \cdot \frac{1-\Phi_n}{\alpha_R + \eta + \delta(k_n/R)^2} \right)}. \quad (9)$$

Hereinafter $\langle \varphi_n(z) \rangle \equiv \Phi_n = \frac{2}{\xi_n h} \cdot \frac{sh(\xi_n h/2)}{ch(\xi_n h/2) + \lambda_b \xi_n sh(\xi_n h/2)}$.

Note, that the expression for paraelectric phase permittivity $\langle \varepsilon_{zz} \rangle$ obtained from Eq.(9) for the infinite cylinder ($h \to \infty$) at $(\lambda_S/R) \to 0$ coincides with one derived by Wang *et. al* [6]. In fact, for the case one can neglect the depolarization field ($\eta \to 0$) and dependence on z ($\varphi_n(z) \to 0$) and the summation on $k_n$ leads to

$$P_Z(\rho, h \to \infty) = E_0 \sum_{n=1}^{\infty} \frac{2J_0(k_n \rho/R)}{k_n J_1(k_n)} \frac{1}{\alpha_R + \delta(k_n/R)^2} \equiv \frac{E_0}{-\alpha_R} \left( \frac{J_0(\rho\sqrt{-\alpha_R/\delta})}{J_0(R\sqrt{-\alpha_R/\delta})} - 1 \right),$$

$$\langle P_Z \rangle = E_0 \sum_{n=1}^{\infty} \frac{4}{k_n^2} \frac{1}{\alpha_R + \delta(k_n/R)^2} \equiv \frac{E_0}{\alpha_R} \left( 1 - \frac{2}{R\sqrt{-\alpha_R/\delta}} \frac{J_1(R\sqrt{-\alpha_R/\delta})}{J_0(R\sqrt{-\alpha_R/\delta})} \right). \quad (10)$$

In order to obtain the solution of Eq. (5), let us use the coordinate-dependent part of the linear solution (8) in the trial function $P_Z(\rho, z) = \sum_{n=1}^{\infty} \frac{2J_0(k_n \rho/R)}{k_n J_1(k_n)} \frac{1-\varphi_n(z)}{1-\Phi_n} P_n$. The variation amplitudes $P_n$ must be determined by the minimization of the expressions (1)-(4). Integration in these expressions with the aforementioned trial function leads to the following form of the free energy:

$$\Delta G = \frac{1}{2} \sum_{n=1}^{\infty} \frac{\alpha_R + \eta + \delta(k_n/R)^2}{1-\Phi_n} \frac{4P_n^2}{k_n^2} - \frac{\eta}{2} \left( \sum_{n=1}^{\infty} \frac{4P_n}{k_n^2} \right)^2 - E_0 \sum_{n=1}^{\infty} \frac{4P_n}{k_n^2} + $$
$$+ \frac{\beta}{4} \left( \sum_{l,n,s,q} g_{lnsq} P_l P_n P_s P_q \right) + \frac{\gamma}{6} \left( \sum_{l,n,s,q,r,t} f_{lnsqrt} P_l P_n P_s P_q P_r P_t \right) \quad (11)$$

Eq. (11) expressed as the algebraic sum of different even powers of polarization components with coefficients dependent on the particles sizes. The exact expressions for the renormalized coefficients in Eq.(11) are rather cumbersome. Note, that hereinafter we suppose that $\xi_n^2 \lambda_b h \gg 1$ and so $\Phi_n \approx 2/h\xi_n(1+\lambda_b\xi_n) \ll 1$, which is typical for the majority of ferroelectric nanoparticles. Really, using typical values $|\alpha| = 10^{-2}$, $\delta = 10^{-19} m^2$ and $\lambda_b = 10^{-8} m$, one can easily obtain that $\xi_n^2 \approx 4\pi/\delta$ and so $\xi_n^2 \lambda_b h \sim 10^2 - 10^3$ for disks of radius $R \gg h$ ($h \sim 10^{-9} m$), whereas $\xi_n^2 \approx ((k_n/R)^2 + \alpha/\delta) \sim 10^{18} m^{-2}$ and thus $\xi_n^2 \lambda_b h \sim 10^1 - 10^2$ for wires of length $h \gg R$ ($R \sim 10^{-9} m$).

The average polarization should be calculated as:

$$\langle P_Z \rangle = \sum_{n=1}^{\infty} \frac{4 P_n}{k_n^2}, \qquad (12)$$

The coupled equations for the amplitudes $P_n$ should be obtained from the variation $\frac{\partial \Delta G}{\partial P_n} = 0$.

The spatial distribution of depolarization field and spontaneous polarization for a single-domain nanorod with $R = 2h$ are depicted in Figs. 2 and 3 respectively.

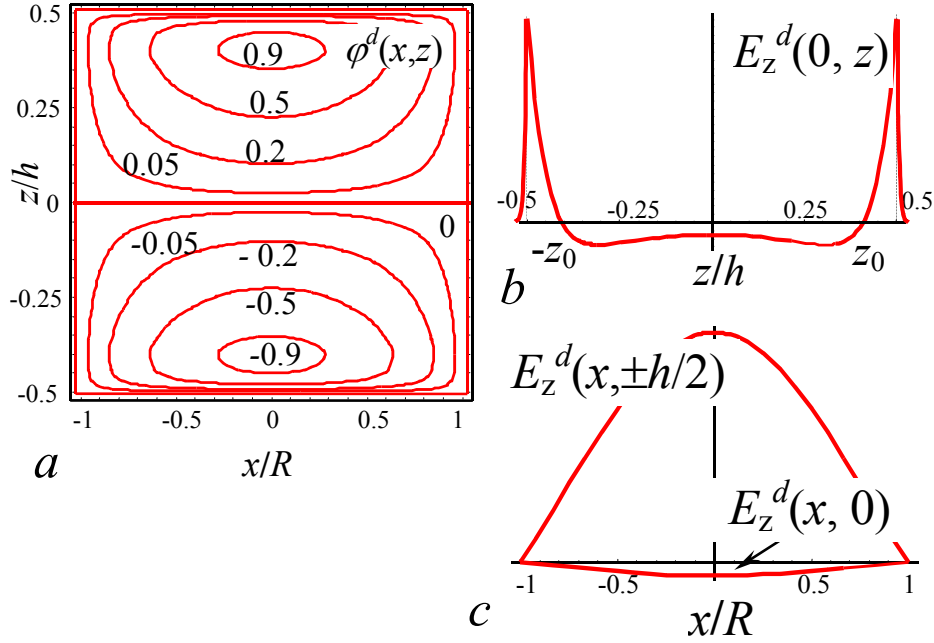

FIG. 2. (Color online) Depolarization field isopotential lines (a), its z-component spatial distribution (b, c) inside the nanorod with $h = 2R$, $\lambda_S \xi = 0.2$, $\xi \approx \sqrt{2\pi/\delta}$. Numbers near the curves in part (a) correspond to the values of potential normalized on its maximal value.

It is seen from Fig.2a that the depolarization field potential $\varphi_z^d(\rho, z)$ is zero on the particle surface allowing for the short-circuit conditions provided by screening charges. Closeness of depolarization field potential $\varphi_z^d(\rho, z)$ contours appears near the faces. It is clear that the gradient (i.e. the depolarization field) is maximal near the cylinder faces $z = \pm h/2$ and minimal in its center $z = 0$ (compare Fig.2a with Fig.2 b,c).

One can see from Fig. 2b that along the nanoparticle axis $\rho = 0$ depolarization field $E_z^d(0, z)$ has maximal positive values near the surfaces $z = \pm h/2$. The depolarization field rapidly decreases inside the particle, reaches zero value at $z = \pm z_0$, then it becomes negative and rather small at $-z_0 \le z \le z_0$. The depolarization field drastically decreases at $|z| \ge h/2$ and completely vanishes outside the screening layer that plays a role of an electrode. It is well known that depolarization field decays exponentially in such conductive layers (see e.g. [17], [27]).

The spatial distribution of depolarization field vs. the distance from nanoparticle axis $\rho = 0$ is depicted in Fig.2c for $z = \pm h/2$ (faces) and $z = 0$ (plane of symmetry), the field being positive and negative respectively. It is clear that the field is maximal at $\rho = 0$ and decreases up to zero when reaching the sidewalls $\rho = R$ as it should be expected for the electric field tangential component on conducting surface.

The behavior of $P_Z(\rho,z)$ for arbitrary $x$ ($y=0$) and $z$ values can be seen from Fig.3a. The spontaneous polarization $P_Z(\rho,z)$ is maximal on the particle axis $\rho = 0$, slightly decreases under approaching its faces $z = \pm h/2$ and essentially decreases under moving away from the polar axis reaching small values on the sidewall surface $\rho = R$ (see values near the contours in Fig.3a).

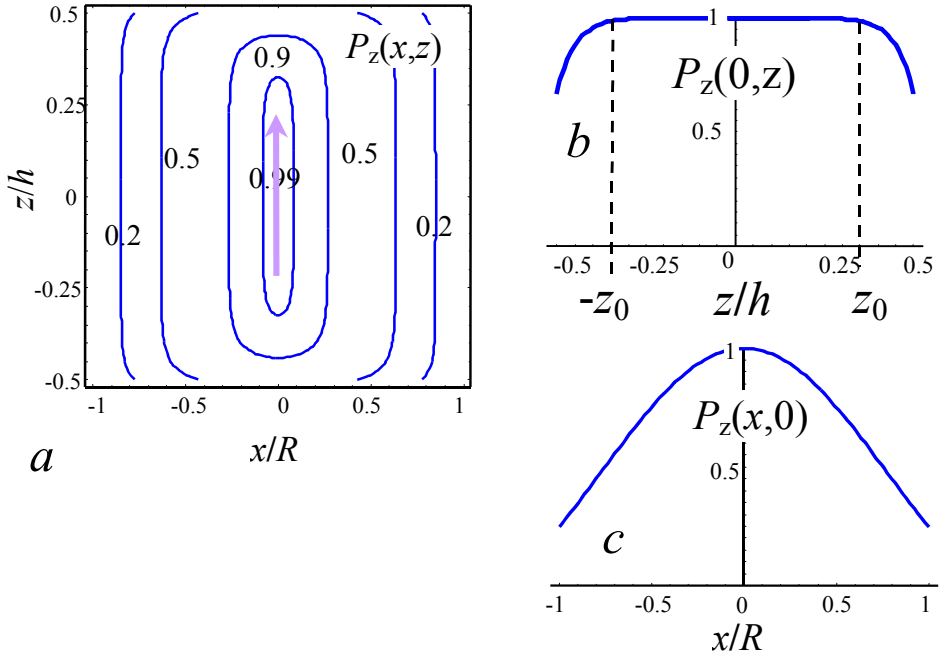

FIG. 3. (Color online) Normalized spontaneous polarization contour lines (a), its spatial distribution (b, c) inside the nanorod with $h = 2R = 10\xi$, $\lambda_b \xi = 2$, $\xi \approx \sqrt{2\pi/\delta}$. Numbers near the contours in part (a) correspond to the values of polarization normalized on its maximal value.

The polarization profile $P_Z(0,z)$ on nanoparticle axis $\rho = 0$ is depicted in Fig. 3b. The polarization is the smallest near the surfaces $z = \pm h/2$, then increases up to the maximal value and flattens in the region $-z_0 \le z \le z_0$.

The behavior of polarization profile $P_Z(\rho,0)$ in the cylinder symmetry plane $z = 0$ is depicted in Fig. 3c. The profile $P_Z(\rho,0)$ is qualitatively similar to that of $P_Z(0,z)$, namely $P_Z(\rho,0)$ has minimal value on the surface $\rho = R$ and reaches the maximum in the central part $\rho = 0$. However, $P_Z(\rho,0)$ profile looks like the diffuse maximum contrary to the plateau shown in Fig. 3b.

The difference of polarization profiles $P_Z(0,z)$ and $P_Z(\rho,0)$ is regarded to the condition $(\lambda_S/R) \ll 1$ and $\xi_n^2 \lambda_b h \gg 1$ used in calculations, as well as related with different functional dependence of $P_Z(\rho,z)$ on $\rho$ (Bessel functions $J_0(k_n \rho/R)$) and on z (hyperbolic functions $1 - \varphi_n(z)$) respectively (see Eqs.(6,7)). The same speculations are valid for the explanation of the difference of depolarization field profiles $E_z^d(0,z)$ and $E_z^d(\rho,0)$, since they are proportional to $\langle P_Z \rangle - P_Z(\rho,z)$ (compare Figs. 2b and 2c).

It is worth to underline, that the profiles and average values of the properties related to spontaneous polarization (e.g. piezoelectric and pyroelectric coefficients) can be calculated with the help of obtained polarization distribution.

## 3. Phase diagram

The equation for paraelectric to ferroelectric phase transition temperature $T_{CN}(R,h)$ can be obtained from Eq. (9) in the following form:

$$1 - \eta(R,h) \sum_{n=1}^{\infty} \frac{4}{k_n^2} \frac{1 - \Phi_n(R,h,T_{CN})}{\alpha_R(T_{CN}) + \eta(R,h) + \delta(k_n/R)^2} = 0. \quad (13)$$

In particular case $\xi_n^2 \lambda_b h \gg 1$ we derived the interpolation for $T_{CN}(R,h)$ valid for the nanodisks, nanorods and nanowires:

$$T_{CN}(R,h) \approx T_C \left(1 - \frac{2\mu Q_{12}}{\alpha_T T_C R} - \frac{k_1^2 \delta}{\alpha_T T_C R^2} - \frac{2\eta(R,h)}{\alpha_T T_C (1+\lambda_b \xi)\xi h}\right), \quad (14)$$

where $\xi(R,h) = \sqrt{\frac{1}{\delta}\left(\frac{k_1^2 \delta}{R^2} + \eta(R,h)\right)}$, $k_1 \approx 2.405$.

The first term in Eq.(14) is the bulk transition temperature, the second term is related to the coupling of surface tension with polarization via electrostriction effect, the third one corresponds to the correlation effects, and the fourth represents the depolarization field contribution. The depolarization field is small enough at $h \gg R$ (see Eq.(3)). The correlation and depolarization terms contribution can only decrease the transition temperature, whereas the second term $\frac{2\mu Q_{12}}{\alpha_T T_C R}$ in Eq. (14) could be positive or negative depending on the $Q_{12}$ sign. Note, that both signs of $Q_{12}$ are possible for different ferroelectrics, however $Q_{12} < 0$ and $Q_{11} + 2Q_{12} > 0$ for most of the perovskite ferroelectrics. Below we demonstrate that increasing of transition temperature and thus ferroelectric properties enhancement and conservation is possible when $\dfrac{2\mu Q_{12}}{\alpha_T T_C R} < 0$ and depolarization field is small enough. This is impossible for the spherical particles with positive value $(Q_{11}+2Q_{12})\mu > 0$ [22].

Let us make some estimation of the second and third terms in Eq.(14) for perovskites BaTiO$_3$ and PbTiO$_3$. Using parameters $Q_{12}=-0.043\,m^4/C^2$, $T_C=400\,K$ (BaTiO$_3$) and $Q_{12}=-0.046\,m^4/C^2$, $T_C=666\,K$ (PbZr$_{0.5}$Ti$_{0.5}$O$_3$) and $\mu=5-50\,N/m$ (see e.g. Ref. 10), $\delta=10^{-19}\,m^2$, we obtained that $\left|\dfrac{2\mu Q_{12}}{\alpha_T T_C}\right|\approx 2-17\,nm$, $\sqrt{\dfrac{\delta k_1^2}{\alpha_T T_C}}\approx 5\,nm$ for BaTiO$_3$ and $\left|\dfrac{2\mu Q_{12}}{\alpha_T T_C}\right|\approx 3-26\,nm$, $\sqrt{\dfrac{\delta k_1^2}{\alpha_T T_C}}\approx 5\,nm$ for PbZr$_{0.5}$Ti$_{0.5}$O$_3$ respectively. So both terms are comparable with unity at nanoparticle radius ~2-25 nm.

For a bulk sample $R\to\infty$, $h\to\infty$ and one obtains that $T_{CN}(R,h)\to T_C$ as it should be expected. For a nanodisk with $R\gg h$ values of depolarization factor $\eta\to 4\pi$ and $\xi_n\to\sqrt{4\pi/\delta}$, so the renormalized transition temperature acquires the form similar to the one derived in Ref. 20. For a nanowire with $h\to\infty$ values $\eta\to 0$ and $\langle\varphi(\xi_n)\rangle\sim 1/\xi_n h\to 0$, i.e. the depolarization field vanishes, thus $T_{CN}(R,h)\approx T_C - \dfrac{2\mu Q_{12}}{\alpha_T R} - \dfrac{k_1^2\delta}{\alpha_T R^2}$. Intermediate situation is realized in prolate nanorods, when the inequality $h\gg R$ is valid, but the small values $\langle\varphi(\xi_n)\rangle\approx \dfrac{2}{(1+\lambda_b\xi_n)\xi_n h}$ and $\eta(R,h)\approx 16\pi\cdot\dfrac{R^2}{h^2}$ are non-zero.

Below we present phase diagrams calculations based on the Eqs. (14). Obtained results are shown in Figs. 4-6. For calculations we introduced the following parameters and dimensionless variables, keeping in mind that $\sqrt{\delta}\sim 0.5\,nm$ is of a lattice constant order:

$$R_\mu = \dfrac{2\mu Q_{12}}{\alpha_T T_C \sqrt{\delta}}, \quad R_S = \sqrt{\dfrac{k_1^2}{\alpha_T T_C}}, \quad r=\dfrac{R}{\sqrt{\delta}}, \quad l=\dfrac{h}{\sqrt{\delta}}, \quad \Lambda_b=\dfrac{\lambda_b}{\sqrt{\delta}}, \quad \Lambda_S=\dfrac{\lambda_S}{\sqrt{\delta}}. \quad (15)$$

In these variables

$$\begin{cases} T_{CN}(r,l)\approx T_C\left(1-\dfrac{R_\mu}{r}-\dfrac{R_S^2}{r^2}-\dfrac{2\eta(r,l)}{\alpha_T T_C(1+\Lambda_b\zeta(r,l))\zeta(r,l)l}\right), \\ \zeta(r,l)=\sqrt{\left(\dfrac{k_1^2}{r^2}+\eta(r,l)\right)}, \qquad k_1\approx 2.405. \end{cases} \quad (16)$$

The critical sizes can be found from the equation $T_{CN}(r,l)=0$. The dependences $l_{cr}(r)$ or $r_{cr}(l)$ determine the boundary between the paraelectric ($l<l_{cr}(r)$ or $r<r_{cr}(l)$) and ferroelectric ($l>l_{cr}(r)$ or $r>r_{cr}(l)$) phase (see solid curves in Fig. 4).

It is clear that ferroelectric phase appreciably broadens at $R_\mu < 0$ in comparison with the one at $R_\mu > 0$. One can see, that at chosen material parameters ferroelectric phase is absent in the nanorods with height $l \leq l_{cr}^{min}$ and arbitrary $r$ ($l_{cr}^{min} \approx 22$ for $R_\mu = +25$ and $l_{cr}^{min} \approx 12$ for $R_\mu = -25$). On the other hand ferroelectric phase is absent in the nanorods with radius $r \leq r_{cr}^{min}$ and arbitrary $l$ ($r_{cr}^{min} \approx 40$ for $R_\mu = +25$ and $r_{cr}^{min} \approx 8$ for $R_\mu = -25$) In both cases we are faced with the size-driven transition from ferroelectric to paraelectric phase (see Fig. 4).

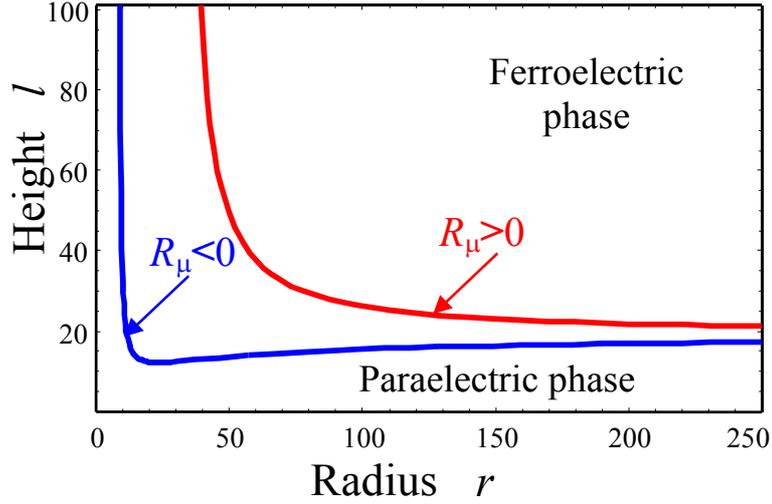

FIG. 4. (Color online) Cylindrical nanoparticles phase diagram. Parameters: $\alpha_T T_C = 3 \cdot 10^{-2}$, $\Lambda_S \leq 1$ and $\Lambda_b = 5$, $R_\mu = \pm 25$, $R_S = 17$. In Figs.4-6 material parameters correspond to PbZr$_{0.5}$Ti$_{0.5}$O$_3$.

The size effect on the phase diagram for the case when the shape of nanoparticle is fixed ($l/2r = const$), but its radius $r$ or length $l$ increases are represented in Figs. 5a, 5b respectively. It is clear from the figures that transition temperature values are different for nanowires ($l/2r \geq 100$), nanorods ($l/2r \geq 10$), nanobars $l/2r \approx 1$ and nanodisks $l/2r \leq 0.1$. The transition temperature tends to the bulk value $T_C$ at $r \to \infty$ and $l \to \infty$ for any shape, as it should be expected for the bulk ferroelectric material.

As it follows from Figs.5a, the transition temperature $T_{CN}(r)$ is the highest for the nanowire, where depolarization field is absent ($\eta \to 0$ for $l \to \infty$), and only the correlation effect and surface tension determine the size dependence of paraelectric-ferroelectric transition temperature. The results are approximately the same for the nanorods with $l/2r \geq 10$. $T_{CN}(r)$ is the lowest for the nanodisks with $l/2r = 0.01$ because of the maximal depolarization factor $\eta \approx 4\pi$.

It is clear from Fig. 5a that the transition temperature $T_{CN}(r)$ between the paraelectric and ferroelectric phase increases monotonically with nanoparticle radius increasing only for $R_\mu > 0$.

For $R_\mu < 0$ nanowires and nanorods reveal the increase of transition temperature up to $T_{CN}^{max}(R_{opt}) = T_C\left(1 + \frac{R_\mu^2}{4R_S^2}\right)$ at radius $R_{opt} = -\frac{2R_S^2}{R_\mu}$ ($T_{CN}^{max}/T_C \approx 1.5$, $R_{opt} \approx 20$). The region of $r$ values where $T_{CN}(r)/T_C > 1$ exists for nanobars, nanorods and nanowires only. No transition temperature increase was obtained for nanodisks ($T_{CN}(r)/T_C$ is always smaller than unity) because of the maximal depolarization factor $\eta \approx 4\pi$. The critical radius is minimal for nanowires and nanorods and increases with $l/2r$ increase.

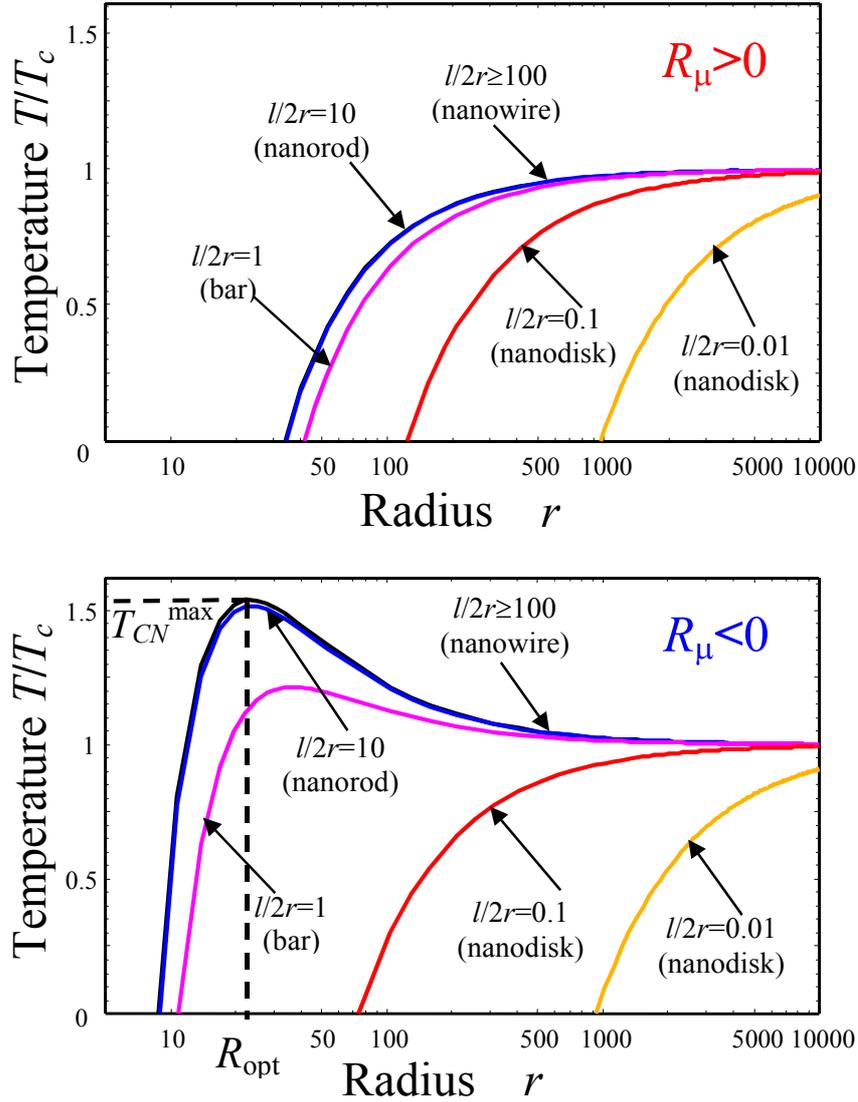

FIG. 5a. (Color online) Transition temperature size dependence for different ratios $l/2r = 100, 10, 1, 0.1, 0.01$. Other parameters: $\alpha_T T_C = 2.8 \cdot 10^{-2}$, $\Lambda_S \leq 1$, $R_\mu = \pm 25$, $R_S = 17$, $\Lambda_b = 5$.

Fig. 5b demonstrates that the transition temperature $T_{CN}(l)$ decreases monotonically with the ratio $l/2r$ increase only for $R_\mu > 0$, namely the nanodisks have the maximal transition temperature, whereas nanowires have the minimal one. For $R_\mu < 0$ nanowires and nanorods reveal the increase of transition temperature up to $T_{CN}^{max} \approx 1.5\, T_C$ at some $l_{opt}$ values. Both for $R_\mu < 0$ and $R_\mu > 0$ the nanodisks have the minimal critical length $l_{cr} \sim 20$, whereas nanowires have the maximal critical length $l_{cr} \sim 2000 - 5000$. These results become clear taking into account that at fixed height $l$ the phase transition in the nanodisk ($2r \gg l$) is influenced minimally by the lateral correlation effects (i.e. the second and the third terms proportional to $1/r$ and $1/r^2$ in Eq.(16) are small in comparison with the ones for the nanorod ($2r \ll l$)).

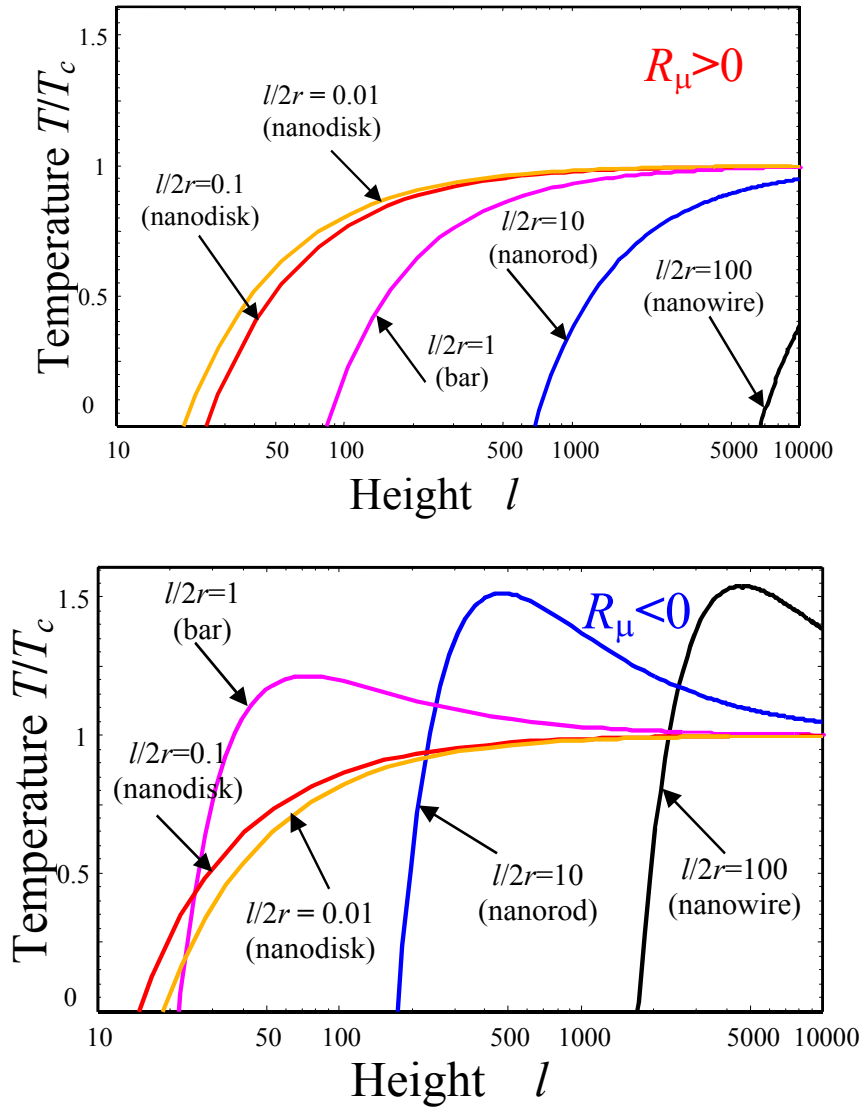

FIG. 5b. (Color online) Transition temperature size dependence for different ratios $l/2r = 100,\ 10,\ 1,\ 0.1,\ 0.01$. Other parameters: $\alpha_T T_C = 2.8 \cdot 10^{-2}$, $\Lambda_S \leq 1$, $R_\mu = \pm 25$, $R_S = 17$, $\Lambda_b = 5$.

It seems natural, that the lateral correlation effect on the size-driven phase transition reveals itself at fixed height, whereas the depolarization field influence on critical sizes and transition temperature is more pronounced at fixed radius (compare curves order for different ratio $l/2r$ in Figs. 5b and 5a).

Finally, let us answer the important question: "What nanoparticle shape posses the minimal critical volume and allows ferroelectricity conservation at higher temperatures?" For this purpose we have to consider the case when the shape of nanoparticle is fixed ($l/2r = const$), but its volume $\pi r^2 l$ increases. The results are represented in Figs. 6.

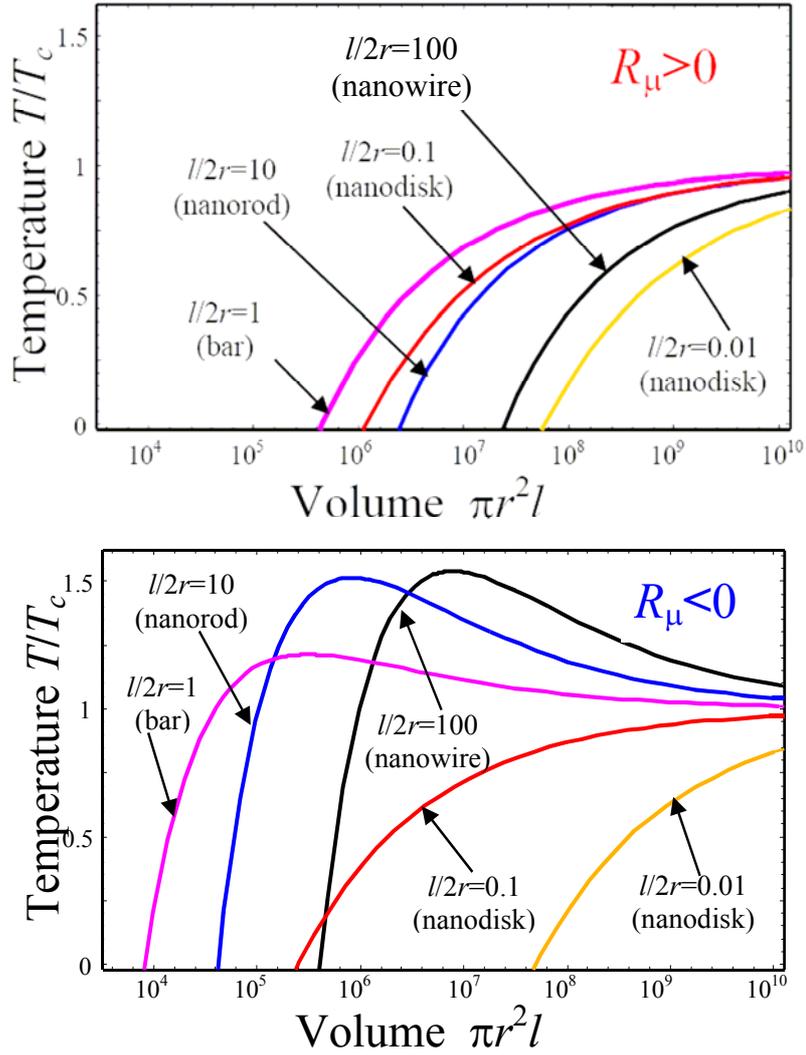

FIG. 6. (Color online) Transition temperature vs. particle volume for different ratios $l/2r = 100, 10, 1, 0.1, 0.01$. Other parameters: $\alpha_T T_C = 2.8 \cdot 10^{-2}$, $\Lambda_S \leq 1$, $R_\mu = \pm 25$, $R_S = 17$, $\Lambda_b = 5$. Note, that the dimensionless volume of $10^{10}$ unit cells corresponds to the 0.64μm³ and to the linear size of 860 nm.

It is follows from the figures that the nanobar ($l/2r = 1$, depolarization factor $\eta \approx 2\pi$) has the smallest critical volume both at $R_\mu > 0$ and at $R_\mu < 0$. Its ferroelectric-paraelectric transition temperature is high enough up to the values higher than $T_C$. Moreover, since $R_\mu \sim Q_{12}\mu < 0$ this nanoparticle shape is preferable in comparison with a spherical one, because a sphere has positive parameter $R_\mu \sim (Q_{11} + 2Q_{12})\mu$ and only slightly smaller depolarization factor $\eta = 4\pi/3$ [21]. The nanorods with the shape $10 < l/2r \leq 100$ reveal the highest transition temperature and thus the best ferroelectric properties in the region of volumes $V > V_{max}$ at $R_\mu < 0$.

Both for $R_\mu > 0$ and $R_\mu < 0$, the dependence of transition temperature on the nanoparticle volume for different shapes $l/2r$ is non-monotonic with respect to the ratio $l/2r$ in contrast to the monotonic radius dependencies (compare Figs. 6 with the Figs. 5a). Let us underline, that under the condition $R_\mu < 0$ nanobars and prolate nanorods ($1 \leq l/2r \leq 10$) posses enhanced polar properties at $r \sim (10 - 200)$ unit cells, namely they have higher spontaneous polarization and transition temperature in comparison with a bulk sample, and the nanobar reveals the minimal critical volume.

## 4. Discussion

We have studied the dependence of ferroelectric nanorod critical sizes on the nanoparticle shape, polarization gradient coefficient, extrapolation length, effective surface tension and electrostriction coefficient $Q_{12}$. Analyzing obtained results we could summarize that finite nanorods of radius $R \sim 20\text{-}200$ unit cells posses the best ferroelectric properties when the electrostriction coefficient $Q_{12}$ is negative, namely they exhibit the higher transition temperature and spontaneous polarization in comparison with the bulk sample, whereas a nanobar ($h/2R = 1$) reveals the minimal critical volume.

Keeping in mind that the condition $R/h \ll 1$ leads to the small values of depolarization field $E_Z^d \sim \eta(R,h) \sim R^2/h^2$, the aforementioned conservation of ferroelectric properties is possible for a sufficiently long nanorod ($h/2R \geq 10$) of small radius $R \sim 2 - 20$ nm. Note, that the existence of poly-domain structure lead to the additional decrease of depolarization field and thus improve the estimation. In the case we obtained, that the increase of $T_{CN}^{max}/T_C > 1$ appears at radius $R_{opt}$ when the parameter $R_\mu \sim Q_{12}\mu$ is negative (see the second term in Eq.(14) and Fig. 5a), i.e. when the effective surface pressure leads to the decrease of the inverse dielectric susceptibility due to the negative electrostriction effect.

Let us qualitatively compare our results with the ones obtained earlier [1], [2], [11] (see Table 1).

**Table 1**

|  | Experiments [1], [11] and *ab initio* calculations [2] | | | Parameters calculated in the framework of proposed model | | | | Material constants used in calculations | | |
| --- | --- | --- | --- | --- | --- | --- | --- | --- | --- | --- |
| Material | $h$, nm | $R_{cr}$, nm | $T_{cr}$, K | $R_{cr}$, nm | $R_{opt}$, nm | $T_{CN}^{max}$, K | $T_C$, K | $Q_{12}$, m$^4$/C$^2$ | $\delta$, $10^{-19}$ m$^2$ | $\mu$, N/m |
| PZT [11] | $\gg R$ | 5-10 | 300 | 1.1-5.3 at 300 K | 2-23 | >704 | 665.7 | -0.046 | 1 | 50-5 |
| BT [2] | $\gg R$ | 1.2 | 0 | 1.1-3.8 at 0 K | 2-24 | >414 | 400 | -0.043 | 1 | 50-5 |
| RS [1] | 500 | ≤30 | 328 | 11-24 at 300 K | 33 | >300 | 297 | $Q_{12}$=1.56 $Q_{13}$=−2.19 [28] | 10 | 5-0.5 |

The applicability of our model to the description of phase transition between cubic paraelectric and tetragonal ferroelectric phases in BT and PZT is for certain. However, it can be only the approximation for improper ferroelastic – ferroelectrics like RS. Therefore our consideration can be applied to experiments of Yadlovker and Berger [1] only in the temperature range where RS ferroelectric properties can be described by the phenomenological expansion (1) over polarization powers.

Obtained results explain ferroelectricity enhancement in the Rochelle salt nanorods [1] of radius $\sim 30\,nm$, piezoelectric properties conservation in lead-zirconate-titanate nanorods [11] of radius $\sim 5-10\,nm$ and are in a good agreement with the first principles calculations in barium titanate nanowires [2] (see Table 1). The predicted effects could be very useful for elaboration of modern nanocomposites with perfect polar properties.

## Appendix A

The free energy expansion on polarization $\mathbf{P}=(0,0,P_3)$ and stress $\sigma_i$ powers has the form [22], [15]:

$$F = a_1 P_3^2 + a_{11} P_3^4 + a_{111} P_3^6 - Q_{11}\sigma_3 P_3^2 - Q_{12}(\sigma_1+\sigma_2)P_3^2 - \\ -\frac{1}{2}s_{11}(\sigma_1^2+\sigma_2^2+\sigma_3^2) - s_{12}(\sigma_1\sigma_2+\sigma_1\sigma_3+\sigma_3\sigma_2) - \frac{1}{2}s_{44}(\sigma_4^2+\sigma_5^2+\sigma_6^2) \quad (A.1)$$

Hereinafter we use Voigt notation $\sigma_i$ or matrix notation $\sigma_{nm}$ (xx=1, yy=2, zz=3, zy=4, zx=5, xy=6) when it necessary.

Firstly let us calculate the $\sigma_i$ components caused by the uniform lateral pressure related to the effective surface tension $p = \mu/R$ [24], [10]. This Lame's problem is discussed in details elsewhere [29]. It is easy to obtain that generalized pressure is directed along the cylinder normal, i.e. $\mathbf{p} \uparrow\uparrow \mathbf{n}$. The conditions of mechanical equilibrium $n_i \sigma_{ij} = -p_j$ on the surface of cylindrical solid body have the following form in the cylindrical coordinates $(\rho, \varphi, z)$:

$$\sigma_{\rho\rho}\big|_{\rho=R} = -p, \quad \sigma_{\rho\varphi}\big|_{\rho=R} = 0, \quad \sigma_{\rho z}\big|_{\rho=R} = 0, \quad \sigma_{zz}\big|_{z=\pm h/2} = 0, \quad \sigma_{z\rho}\big|_{z=\pm h/2} = 0, \quad \sigma_{z\varphi}\big|_{z=\pm h/2} = 0 \tag{A.2}$$

The conditions of mechanical equilibrium $\partial \sigma_{ij}/\partial x_i = 0$ in the bulk of solid body are the following:

$$\begin{cases} \dfrac{\partial \sigma_{zz}}{\partial z} + \dfrac{\partial \sigma_{z\rho}}{\partial \rho} + \dfrac{\sigma_{z\rho}}{\rho} + \dfrac{1}{\rho}\dfrac{\partial \sigma_{z\varphi}}{\partial \varphi} = 0, \\ \dfrac{\partial \sigma_{\rho\rho}}{\partial \rho} + \dfrac{\sigma_{\rho\rho} - \sigma_{\varphi\varphi}}{\rho} + \dfrac{1}{\rho}\dfrac{\partial \sigma_{\rho\varphi}}{\partial \varphi} + \dfrac{\partial \sigma_{z\rho}}{\partial z} = 0, \\ \dfrac{1}{\rho}\dfrac{\partial \sigma_{\varphi\varphi}}{\partial \varphi} + \dfrac{\partial \sigma_{z\varphi}}{\partial z} + \dfrac{\partial \sigma_{\rho\varphi}}{\partial \rho} + 2\dfrac{\sigma_{\rho\varphi}}{\rho} = 0. \end{cases} \tag{A.3}$$

It is seen that boundary and equilibrium conditions (A.2) and (A.3) can be fulfilled with

$$\sigma_{\rho\rho} = \sigma_{\varphi\varphi} = -p, \quad \sigma_{\rho\varphi} = 0, \quad \sigma_{\rho z} = 0, \quad \sigma_{zz} = 0, \quad \sigma_{z\varphi} = 0. \tag{A.4}$$

The tensor components in Cartesian coordinates can be found from relations:

$$\begin{pmatrix} \sigma_{xx} & \sigma_{xy} & \sigma_{xz} \\ \sigma_{yx} & \sigma_{yy} & \sigma_{yz} \\ \sigma_{zx} & \sigma_{zy} & \sigma_{zz} \end{pmatrix} = \begin{pmatrix} \cos(\varphi) & -\sin(\varphi) & 0 \\ \sin(\varphi) & \cos(\varphi) & 0 \\ 0 & 0 & 1 \end{pmatrix} \cdot \begin{pmatrix} \sigma_{\rho\rho} & \sigma_{\rho\varphi} & \sigma_{\rho z} \\ \sigma_{\varphi\rho} & \sigma_{\varphi\varphi} & \sigma_{\varphi z} \\ \sigma_{z\rho} & \sigma_{z\varphi} & \sigma_{zz} \end{pmatrix} \cdot \begin{pmatrix} \cos(\varphi) & \sin(\varphi) & 0 \\ -\sin(\varphi) & \cos(\varphi) & 0 \\ 0 & 0 & 1 \end{pmatrix} \tag{A.5}$$

Allowing for Eq. (A.4), expression (A.5) leads to

$$\begin{aligned} \sigma_{xx} &= \cos(\varphi)^2 \sigma_{\rho\rho} + \sin(\varphi)^2 \sigma_{\varphi\varphi} = -p, \\ \sigma_{yy} &= \sin(\varphi)^2 \sigma_{\rho\rho} + \cos(\varphi)^2 \sigma_{\varphi\varphi} = -p, \\ \sigma_{xy} &= \cos(\varphi)\sin(\varphi)(\sigma_{\rho\rho} - \sigma_{\varphi\varphi}) = 0, \\ \sigma_{xz} &= \sigma_{yz} = \sigma_{zz} = 0. \end{aligned} \tag{A.6}$$

In Voigt notation this gives:

$$\sigma_1 = \sigma_2 = -\frac{\mu}{R}, \qquad \sigma_3 = \sigma_4 = \sigma_5 = \sigma_6 = 0. \tag{A.7}$$

The anzats of solutions (A.7) into the free energy (A.1) gives the expression:

$$F = \left(a_1 + 2Q_{12}\frac{\mu}{R}\right)P_3^2 + a_{11}P_3^4 + a_{111}P_3^6 - (s_{11} + s_{12})\frac{\mu^2}{R^2} \tag{A.8}$$

The minimization of free energy (A.8) on the polarization components $\partial F/\partial P_3 = E_0$ gives the equation of state.

Note, that the renormalization of coefficient $a_1^* = (a_1 + 2Q_{12}p)$ for a cylindrical nanoparticle differs from the one $a_1^* = (a_1 + (Q_{11} + 2Q_{12})p)$ obtained for a spherical nanoparticle recently [22]. Both results are clear owing to the fact that $\sigma_1 = \sigma_2 = -p$ and $\sigma_3 = 0$ for a cylinder, whereas $\sigma_1 = \sigma_2 = \sigma_3 = -p$ for a sphere. Also let us underline that we do not take into account possible stress relaxation caused by dislocations. This approach used by many authors (see e.g. Refs. 20, 15) is valid under the conditions discussed elsewhere [25].

**Appendix B**

Let us consider the depolarization field distribution for the case of cylindrical particle with arbitrary polarization distribution in the ambient conditions. In the equilibrium the perfect screening can be achieved so that there will be no electric field outside the particle.

The field distribution can be obtained on the basis of the electrostatic Poisson's equation for the electric potential $\varphi$:

$$\Delta\varphi(\rho, z) = 4\pi \, \text{div}\, \mathbf{P}(\rho, z) \tag{B.1}$$

Here $\mathbf{P}(\rho, z) = (0, 0, P_Z)$ is the given z-component polarization distribution inside the particle, which has the cylindrical symmetry: $\Delta = \dfrac{\partial^2}{\partial z^2} + \dfrac{1}{\rho}\dfrac{\partial}{\partial \rho}\rho\dfrac{\partial}{\partial \rho}$. The boundary conditions on the particle surface has the view:

$$\varphi(\rho = R, z) = 0, \quad \varphi\left(\rho, z = -\frac{h}{2}\right) = 0, \quad \varphi\left(\rho, z = \frac{h}{2}\right) = U. \tag{B.2}$$

Here $R$ and $h$ is the cylinder radius and height respectively, $U$ is the applied voltage. At $U = 0$ the boundary conditions (B.2) corresponds to the short-circuit ones proposed by Kretschmer and Binder [19] for a film.

The system (B.1), (B.2) can be solved by means of the separation of variables method. Since for the system of cylindrical symmetry eigen-functions of Laplace operator $\Delta$ are the Bessel functions one can find the potential $\varphi$ in the form of series:

$$\varphi(\rho, z) = \sum_{n=1}^{\infty} C_n(z) J_0\left(k_n \frac{\rho}{R}\right) \tag{B.3}$$

Here $J_0(x)$ is the first kind Bessel function of zero order, $k_n$ is the n-th root of this function ($J_0(k_n) = 0$) and functions $C_n(z)$ should satisfy the following boundary problem:

$$\begin{cases} \dfrac{d^2 C_n(z)}{dz^2} - \left(\dfrac{k_n}{R}\right)^2 C_n(z) = 4\pi \dfrac{2}{R^2 J_1(k_n)^2} \int_0^R \dfrac{\partial P_Z(\rho, z)}{\partial z} J_0\left(k_n \dfrac{\rho}{R}\right) \rho\, d\rho \\ C_n\left(z = -\dfrac{h}{2}\right) = 0, \quad C_n\left(z = \dfrac{h}{2}\right) = \dfrac{2}{k_n J_1(k_n)} U \end{cases} \tag{B.4}$$

In (B.4) we used the Bessel functions orthogonality $\int_0^R J_0\left(k_n \frac{\rho}{R}\right) J_0\left(k_m \frac{\rho}{R}\right) \rho \, d\rho = \delta_{nm} \frac{R^2 J_1(k_n)^2}{2}$, integral $\int_0^R J_0\left(k_n \frac{\rho}{R}\right) \rho \, d\rho = R^2 \frac{J_1(k_n)}{k_n}$ and expansion $1 = \sum_{n=1}^\infty \frac{2}{k_n J_1(k_n)} J_0\left(k_n \frac{\rho}{R}\right)$ at $\rho < R$. In accordance with the general theory of the linear second order differential equations one can find the solution of (B.4) in the form:

$$C_n(z) = \sum_{m=1}^\infty g_{mn} \sin\left(\frac{2\pi m z}{h}\right) + A_n \exp\left(\frac{k_n z}{R}\right) + B_n \exp\left(-\frac{k_n z}{R}\right),$$

$$g_{mn} = -\frac{4\pi}{\left(\frac{2\pi m}{h}\right)^2 + \left(\frac{k_n}{R}\right)^2} \int_{-h/2}^{h/2} \frac{2 dz}{h} \sin\left(\frac{2\pi m z}{h}\right) \int_0^R \frac{2\rho \, d\rho}{R^2 J_1(k_n)^2} J_0\left(k_n \frac{\rho}{R}\right) \frac{\partial P_Z(\rho, z)}{\partial z},$$  (B.5a)

$$A_n = \frac{2U}{k_n J_1(k_n)} \frac{\exp\left(-\frac{k_n h}{2R}\right)}{1 - \exp\left(-\frac{2 k_n h}{R}\right)}, \quad B_n = -\frac{2U}{k_n J_1(k_n)} \frac{\exp\left(-\frac{3 k_n h}{2R}\right)}{1 - \exp\left(-\frac{2 k_n h}{R}\right)}.$$ (B.5b)

Keeping in mind (B.3) and (B.5) one obtains that depolarization field z-component $E_Z^d = -\partial \varphi(\rho, z)/\partial z$ acquires the form after integrating over parts:

$$E_Z^d(\rho, z) = \sum_{m=1}^\infty \sum_{n=1}^\infty E_{mn} \cos\left(\frac{2\pi m}{h} z\right) J_0\left(k_n \frac{\rho}{R}\right),$$

$$E_{mn} = -4\pi \frac{(2\pi m R)^2}{(2\pi m R)^2 + (k_n h)^2} P_{mn},$$ (B.6)

$$P_{mn} = \int_{-h/2}^{h/2} \frac{2 d\tilde{z}}{h} \cos\left(\frac{2\pi m}{h} \tilde{z}\right) \int_0^R \frac{2\tilde{\rho} \, d\tilde{\rho}}{R^2 J_1(k_n)^2} J_0\left(k_n \frac{\tilde{\rho}}{R}\right) P_Z(\tilde{\rho}, \tilde{z}).$$

Note, that coefficients $P_{mn}$ coincide with the ones in polarization expansion:

$$P_Z(\rho, z) = \sum_{m=0}^\infty \sum_{n=1}^\infty P_{mn} \cos\left(\frac{2\pi m}{h} z\right) J_0\left(k_n \frac{\rho}{R}\right).$$ (B.7)

It should be noticed that contrast to (B.6) the expansion (B.7) contains the terms with $P_{0n} = \int_{-h/2}^{h/2} \frac{2 d\tilde{z}}{h} \int_0^R \frac{2\tilde{\rho} \, d\tilde{\rho}}{R^2 J_1(k_n)^2} J_0\left(k_n \frac{\tilde{\rho}}{R}\right) P_Z(\tilde{\rho}, \tilde{z})$ related to the average polarization $\langle P_Z(\rho, z) \rangle = \sum_{n=1}^\infty P_{0n} \frac{2 J_1(k_n)}{k_n}$. The difference $P_Z(\rho, z) - \langle P_Z(\rho, z) \rangle$ acquires the form:

$$P_Z(\rho, z) - \langle P_Z(\rho, z) \rangle = \sum_{m=1}^\infty \sum_{n=1}^\infty P_{mn} \cos\left(\frac{2\pi m}{h} z\right) J_0\left(k_n \frac{\rho}{R}\right) + \sum_{n=1}^\infty \left(J_0\left(k_n \frac{\rho}{R}\right) - \frac{2 J_1(k_n)}{k_n}\right) P_{0n}$$ (B.8)

Let us assume the good convergence of the series in (A.7)-(A.8) and consider the particular following cases.

1) In the particular case $h \ll \pi R$ one obtains that $\dfrac{1}{1+(k_n h/2\pi m R)^2} \approx 1$ at $n \leq m$ and

$P_{mn} \approx \dfrac{2}{k_n J_1(k_n)} \int\limits_{-h/2}^{h/2} \dfrac{2d\tilde{z}}{h} \cos\left(\dfrac{2\pi m}{h}\tilde{z}\right) P_Z(0,\tilde{z})$ in accordance with Laplace method. Then we derive that the term $\sum\limits_{n=1}^{\infty}\left(J_0\left(k_n\dfrac{\rho}{R}\right) - \dfrac{2J_1(k_n)}{k_n}\right) P_{0n} = 0$ allowing for the unity expansion $1 = \sum\limits_{n=1}^{\infty} \dfrac{2}{k_n J_1(k_n)} J_0\left(k_n\dfrac{\rho}{R}\right)$ and equalities $1 \equiv \sum\limits_{n=1}^{\infty} \dfrac{4}{k_n^2} \equiv \sum\limits_{n=1}^{\infty} \dfrac{2}{k_n J_1(k_n)}$, which can be easily obtained from the unity expansion. Thus the

approximate expression for depolarization field has the form (see (B.6)-(B.8)):

$$E_Z^d(z) \approx -4\pi\left(P_Z(z) - \langle P_Z(z)\rangle\right) \tag{B.9}$$

Note, that (B.9) is exact at $R \to \infty$ and coincides with the one obtained for ferroelectric films [19] at $P_Z(\rho,z) \equiv P_Z(z)$.

2) In the particular case $h \gg \pi R$ one obtains that $\dfrac{1}{1+(k_n h/2\pi m R)^2} \approx \left(\dfrac{2R}{h}\right)^2$ at $n \leq m$. Thus the

depolarization field is rather small in comparison with (B.9), namely:

$$\left|E_Z^d(\rho,z)\right| \sim 4\pi\left(\dfrac{2R}{h}\right)^2 \left(P_{Z^2}(z) - \langle P_Z(z)\rangle\right). \tag{B.10}$$

The interpolation for the depolarization field that contains the aforementioned particular cases (B.9)-(B.10) acquires the form:

$$E_Z^d(\rho,z) = -\dfrac{4\pi}{1+(h/2R)^2}\left(P_Z(z) - \langle P_Z(z)\rangle\right) \tag{B.11}$$

Note, that Yadlovker and Berger [1] observed ferroelectric domains with walls oriented along the rod polar axis in a nanoparticle of RS. Keeping in mind, that the domain wall energy are represented by the correlation term $\dfrac{\delta}{2}(\nabla P_Z(\rho,z))^2$ in Eq.(1) for the continuous media approximation, polydomain states could be studied with the help of the free energy (1)-(4). However, for these states adequate description one should use exact expression (B.6) for the depolarization field and calculate the polarization distribution $P_Z(\rho,z) = \sum\limits_{m=0}^{\infty}\sum\limits_{n=1}^{\infty} P_{mn} \cos\left(\dfrac{2\pi m}{h}z\right) J_0\left(k_n\dfrac{\rho}{R}\right)$ in accordance with Eqs. (B.7), (B.6) and (5) self-consistently. The contour lines of depolarization field isopotential lines and spontaneous polarization spatial distribution for nanorod with $R = 2h$ containing one cylindrical domain are depicted in Fig. 7.

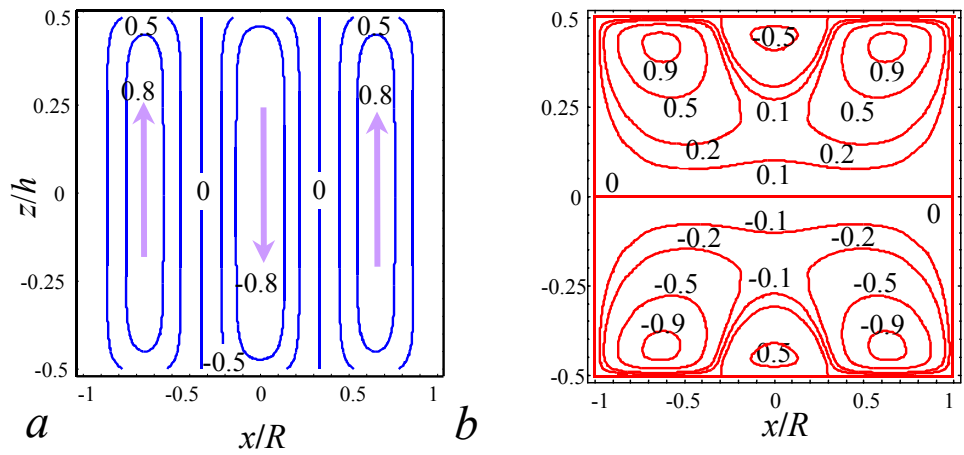

FIG. 7. (Color online) Depolarization field isopotential lines (a) and contour lines of spontaneous polarization spatial distribution (b) for poly-domain state of the nanorod with $h = 2R$. Numbers near curves correspond to the values of polarization and potential normalized on their maximal values.